\documentclass[a4paper,11pt]{article}

\usepackage{jheppub} 

\usepackage{epsfig,multicol}

\newcommand\fverb{\setbox\pippobox=\hbox\bgroup\verb}
\newcommand\fverbdo{\egroup\medskip\noindent%
            \fbox{\unhbox\pippobox}\ }
\newcommand\fverbit{\egroup\item[\fbox{\unhbox\pippobox}]}
\newbox\pippobox
\title{One-Loop Effective Action and Schwinger Effect in (Anti-) de Sitter Space}

\author[a,1]{Rong-Gen Cai}
\author[b,c,2]{Sang Pyo Kim}

\affiliation[a]{State Key Laboratory Theoretical Physics,
Institute of Theoretical Physics, Chinese Academy of Sciences,\\
Beijing 100190, China}
\affiliation[b]{Department of Physics, Kunsan National University,\\ Kunsan 573-701,
 Korea}
\affiliation[c]{Kavli Institute of Theoretical Physics China, Chinese Academy of Sciences,\\
Beijing 100190, China}

\emailAdd{cairg@itp.ac.cn}
\emailAdd{sangkim@kunsan.ac.kr}

\preprint{}  

\abstract{We study the Schwinger mechanism by a uniform electric field in ${\rm dS}_2$ and ${\rm AdS}_2$ and the curvature effect on the Schwinger effect, and further propose a thermal interpretation of the Schwinger formula in terms of the Gibbons-Hawking temperature and the Unruh temperature for an accelerating charge in ${\rm dS}_2$ and an analogous expression in ${\rm AdS}_2$. The exact one-loop effective action is found in the proper-time integral in each space, which is determined by the effective mass, the Maxwell scalar, and the scalar curvature, and whose pole structure gives the imaginary part of the effective action and the exact pair-production rate. The exact pair-production rate is also given the thermal interpretation.}


\begin{document}
\maketitle
\flushbottom

\section{Introduction}

A strong electric field through the electromagnetic interaction with virtual pairs of the vacuum produces charged real pairs,
known as the Schwimger mechanism, and polarizes the vacuum and gives nonlinear quantum electrodynamics (QED) action \cite{Schwinger}. A de Sitter (dS) space is known to produce the Gibbons-Hawking radiation from the cosmological horizon \cite{Gibbons-Hawking}, around which one species of virtual pairs is emitted to an observer and the other species falls behind the horizon, implying the dS radiation \cite{Parikh-Wilczek}. One interesting question is then what effect the spacetime curvature has on the pair production from a uniform electric field and conversely how the electric field affects the dS radiation. A more challenging task will be to find the one-loop effective action both in the uniform electric field and in the curved spacetimes. The exact one-loop effective action in a general electromagnetic field or a general spacetime, however, has not been found yet in the sense of the Schwinger action in the proper-time integral for a uniform electromagnetic field though the one-loop effective action can be formally written in the heat kernel method \cite{DeWitt03}.  Because the pair production and the vacuum polarization are nonperturbative QED phenomena, the one-loop effective action is essential in exploring the quantum structure of the vacuum  as well as the pair production.

The dS space is an interesting issue both in cosmology and black hole physics. Remarkably, the Friedmann-Robertson-Walker equation for the dS space can be explained by the first law of black hole thermodynamics \cite{Cai-Kim}. The vacuum of the dS space is important in understanding the early universe. The dS space in the global coordinates has an intriguing property of vanishing dS radiation in odd spacetime dimensions \cite{Bousso-Maloney-Strominger}. The stability of odd-dimensional dS spaces and the instability of even-dimensional dS spaces may be interpreted as the solitonic nature \cite{Polyakov} or the Stokes phenomenon \cite{Kim10,Kim13a}. QED might have played some role in the early stage of the universe though the electromagnetic field in the present universe is not strong enough to polarize the vacuum and to create pairs. The radial motion of a scalar wave in the Nariai geometry of a rotating black hole in the dS space is equivalent to a massive charge in a uniform electric field in the ${\rm dS}_2$ space \cite{Anninos-Hartman10,Anninos-Anous10}. The near-horizon geometry of an extremal rotating black hole or an extremal Reissner-Nordstr\"{o}m (RN)  black hole has the topology ${\rm AdS}_2 \times {\rm S}^2$ and the motion of a charged scalar field is equivalent to that in the uniform electric field in the ${\rm AdS}_2$ space \cite{Bardeen-Horowitz}. Thus, the Schwinger mechanism in ${\rm AdS}_2$ may shed light on understanding the Hawking radiation near the extremal rotating black hole and the pair production by the near extremal RN black hole \cite{CKLSW}.

The dS spaces with the maximal spacetime symmetry allow explicit solutions of a charged field in the uniform electric field and thereby lead to the exact pair-production rate \cite{Garriga,Villalba,Mendy,Kim-Page08,BCD,GKSSTV,Haouat-Chekireb,FGKSSTV,Kim14}. Similarly, the quantum states of a charged scalar in the uniform electric field in ${\rm AdS}_2$ gives the Breitenlohler-Freedman bound for the pair production, a condition for violating the stability of AdS space \cite{Kim-Page08,Pioline-Troost}. The dS radiation of a scalar field has a Bose-Einstein distribution with the Gibbons-Hawking temperature while the Schwinger formula for the pair production in the uniform electric field has a Boltzmann distribution, regardless of the spin of particles. Furthermore, charged particles or virtual pairs in the uniform electric field could accelerate for an infinite period and feel a thermal spectrum with the Unruh temperature due to the Rindler horizon \cite{Davies,Unruh}. The worldline instanton path of a charge in the unform electric field undergoes a periodic motion in the proper time and thus suggests the Unruh temperature to be associated with the non-relativistic acceleration \cite{Dunne-Schubert}. Can one then understand the spectrum of pairs produced by the uniform electric field in the dS space as that of accelerating charges? Interestingly, the spectrum felt by an accelerating observer in the dS space has the effective temperature, which is the geometric mean with the exponent two of the Unruh temperature and the Gibbons-Hawking temperature \cite{NPT,Deser-Levin97}. The constant curvature black hole also has an effective temperature for the Hawking radiation \cite{Cai-Myung}.

The purpose of this paper is two-fold. First, we propose the effective temperature interpretation of the Schwinger pair production both in a uniform electric field and in the ${\rm dS}_2$ and ${\rm AdS}_2$ space. We then find the exact one-loop effective actions and discuss physical implications to black holes and gravity. The Schwinger formula given by the Boltzmann factor, the leading term of the exact pair-production rate, can be found from the instanton action in the tunneling picture \cite{Kim-Page07,Dumlu-Dunne}. The instanton action is determined both by the scalar curvature of the dS or AdS space and by the acceleration of charge by the electric field. We observe that the Schwinger formula may be interpreted by the effective temperature analogous to that seen by an accelerating observer in the given space except for a factor of two, which is intrinsic to the Schwinger mechanism in the Minkowski spacetime. The exact pair-production rate from the Bogoliubov transformation of the quantized charged field, however, has an additional temperature, which is also expressed by the Gibbons-Hawking and the Unruh temperatures, and which becomes significant only when the Gibbons-Hawking temperature or the scalar curvature is larger than the Unruh temperature.

In order to find the exact one-loop effective action, we employ the in-out formalism by Schwinger and DeWitt \cite{DeWitt03,DeWitt75}, in which the one-loop action is the scattering matrix between the in-vacuum and the out-vacuum, which in turn can be expressed by the Bogoliubov coefficient. For that purpose, we find the Bogoliubov coefficient of a charged scalar field in the uniform electric field and in ${\rm dS}_2$ and ${\rm AdS}_2$, and then use the gamma-function regularization method, which has recently been introduced to compute QED actions in time-dependent or spatially localized electric fields \cite{Kim-Lee-Yoon08,Kim-Lee-Yoon10}. The Bogoliubov coefficient of the charged field is given by a fraction of gamma functions and each gamma function has a proper-time integral representation, leading to a renormalized complex action for the pair production either by the electric field or by ${\rm dS}_2$. We investigate both the effect of curvature on the Schwinger mechanism, which corresponds to the weak gravity limit, and the effect of the Schwinger mechanism on the dS radiation, which corresponds to the weak field limit. We find the power series expansion of the one-loop effective action in terms of the scalar curvature and the Maxwell scalar in ${\rm dS}_2$ and ${\rm AdS}_2$.

The organization of this paper is as follows. In section 2, we find the relativistic (WKB) instanton actions for a charged scalar in a constant electric field in ${\rm dS}_2$ and ${\rm AdS}_2$, and interpret the Schwinger pair production in terms of an effective temperature analogous to the accelerating observer in the given space. In section 3, we quantize the charged scalar field in ${\rm dS}_2$ and find the pair-production rate through the Bogoliubov transformation. We further find a complex one-loop effective action in the in-out formalism, whose real part is the vacuum polarization, and whose imaginary part is the vacuum persistence. We show that the one-loop effective action satisfies the consistence relation between the pair-production rate and the vacuum persistence. We discuss the curvature effect on the Schwinger mechanism and also the Schwinger effect on the dS radiation. In section 4, we quantize the charged field in ${\rm AdS}_2$, find the Bogoluibov coefficient from the tunneling boundary condition, and compute the exact one-loop effective action. We discuss the Breitenlohler-Freedman bound and the curvature effect on the Schwinger mechanism. Finally, we discuss the physical implication of the result in section 5.

\section{Effective Temperature for Schwinger Effect  in ${\rm dS}_2$ and ${\rm AdS}_2$}

In a curved spacetime with the metric tensor $g_{\mu \nu}$, a charged scalar field with the mass $m$ and the charge $q$ obeys the equation (units of $c = \hbar = k_B =1$)
\begin{eqnarray}
\frac{1}{\sqrt{-g}} {\cal D}_{\mu} \Bigl(\sqrt{-g} g^{\mu \nu} {\cal D}_{\nu} \Bigr) \phi - m^2 \phi = 0,
\end{eqnarray}
where ${\cal D}_{\mu} = \partial_{\mu} - i q A_{\mu}$ with $A_{\mu} = (- A_0, A_{i})$ and the field tensor is $F_{\mu \nu} = A_{\nu, \mu} - A_{\mu, \nu}$ in the curved spacetime. In the planar coordinates of ${\rm dS}_2$, a constant electric field has the vector potential
\begin{eqnarray}
ds^2 = -dt^2 + e^{2Ht}dx^2, \quad A_1 (t) = - \frac{E}{H} (e^{Ht} -1), \label{ds gauge}
\end{eqnarray}
while in ${\rm AdS}_2$ space it has the Coulomb potential
\begin{eqnarray}
ds^2 = - e^{2Kx} dt^2 + dx^2, \quad A_0 (t) = - \frac{E}{K} (e^{Kx} -1). \label{ads gauge}
\end{eqnarray}
The gauge potentials reduce to those in the Minkowski space in the limit of $H = 0$ or $K=0$.
The scalar curvature and the Maxwell scalar are, respectively,
\begin{eqnarray}
{\cal R}_{dS} = 2 H^2, \quad {\cal R}_{AdS} = - 2 K^2, \quad {\cal F} = \frac{1}{4} F_{\mu \nu} F^{\mu \nu} = - \frac{1}{2} E^2.
\end{eqnarray}

In ${\rm dS}_2$, the momentum mode $\phi_{k} = e^{-Ht/2} \varphi_{k}$  has the equation
\begin{eqnarray}
\ddot{\varphi}_{k} (t) + \bar{\Omega}_{k}^2 (t) \varphi_{\bf k} (t) = 0, \label{ds mod}
\end{eqnarray}
where the dot denotes the derivative with respect to time and
\begin{eqnarray}
\bar{\Omega}_{k}^2 (t) = \gamma_{dS}^2 + \bar{k}^2 e^{-2Ht} + 2 \frac{qE}{H} \bar{k} e^{-Ht}, \quad \gamma_{dS}^2 =  \Bigl(\frac{qE}{H} \Bigr)^2 + m^2  -  \frac{H^2}{4}, \label{ds freq}
\end{eqnarray}
and $\bar{k} = k - qE/H$ is the shifted momentum. Note that $\gamma_{dS}$ is the frequency of the charge with the momentum $k$ near the future infinity or the cosmological horizon.
In ${\rm AdS}_2$, the energy mode $\phi_{\omega} = e^{-Kx/2} \varphi_{\omega}$
now has the equation
\begin{eqnarray}
{\varphi}''_{\omega} (x) + \bar{\Omega}_{\omega}^2 (x) \varphi_{\omega} (x) = 0, \label{ads mod}
\end{eqnarray}
where the prime denotes the derivative with respect to space and
\begin{eqnarray}
\bar{\Omega}_{\omega}^2 (x) = \gamma_{AdS}^2 + \bar{\omega}^2 e^{-2Kx} - 2 \frac{qE}{K} \bar{\omega} e^{-Kx}, \quad \gamma^2_{AdS} = \Bigl(\frac{qE}{K} \Bigr)^2 - \Bigl(  m^2 + \frac{K^2}{4} \Bigr) \label{ads freq}
\end{eqnarray}
and $\bar{\omega} = \omega - qE/K$ is the shifted energy. The charge with the energy $\omega$ has the real momentum $\gamma_{AdS}$ near the asymptotic boundary. The positive energy solution to eq. (\ref{ds mod}) may have the WKB form $\varphi_k = e^{- iS_k(t)}$ and the positive flux solution to eq. (\ref{ads mod}) may have another WKB form $\varphi_{\omega} = e^{iS_{\omega} (x)}$.
Then, in the tunneling picture, the relativistic instanton action in ${\rm dS}_2$ is given by the phase-integral \cite{Kim-Page07,Dumlu-Dunne}, which from the formula 2.266 of ref. \cite{GR-table} leads to
\begin{eqnarray}
S_k = \int_{t_{c}}^{t_c^*} \bar{\Omega}_k (t) dt = - i \pi \Bigl( \frac{\gamma_{dS}}{H} - \frac{qE}{H^2} \Bigr), \label{ds ins}
\end{eqnarray}
where $t_{c}$ and $t_c^*$ are a pair of complex turning points in the complex time. Similarly, the instanton action in ${\rm AdS}_2$ is by
\begin{eqnarray}
S_{\omega} = \int_{x_{c}}^{x_c^*} \bar{\Omega}_{\omega} (x) dx = - i \pi \Bigl( \frac{\gamma_{AdS}}{K} - \frac{qE}{K^2} \Bigr), \label{ads ins}
\end{eqnarray}
where $x_c$ and $x_c^*$ are another pair of complex turning points in the complex space. We have used the translational- or time-symmetric argument in obtaining eqs. (\ref{ds ins}) and (\ref{ads ins}).
Another complex method gives the pair-production rate by the contour integral as ${\rm Im} S_{k} = (1/2) \oint \bar{\Omega}_{k} (z) dz$ and ${\rm Im} S_{\omega} = - (1/2) \oint \bar{\Omega}_{\omega} (z) dz$ \cite{Kim13a,Kim14,Kim-Page07,Kim13b}. Then, the Schwinger formula for pairs with the given momentum or energy is approximately given by
\begin{eqnarray}
N_{\rm S} = e^{- 2{\cal S}}, \label{pair}
\end{eqnarray}
where the instanton action for ${\rm dS}_2$ is
\begin{eqnarray}
{\cal S}_{k} = - 2~{\rm Im} S_{k} = \frac{2 \pi}{H} \Bigl( \gamma_{dS} - \frac{qE}{H} \Bigr)
\end{eqnarray}
and for and ${\rm AdS}_2$ is
\begin{eqnarray}
{\cal S}_{\omega} =  2~{\rm Im} S_{\omega} = \frac{2 \pi}{K}\Bigl(\frac{qE}{K}- \gamma_{AdS} \Bigr).
\end{eqnarray}
The pair-production rate (\ref{pair}) is the leading Boltzmann factor of the exact formula in sections 3 and 4 below.

In the pure ${\rm dS}_2$ ($E=0$), a species of particles with the mass $m$ can be produced provided that the Compton wavelength $\lambda_{\rm C} = 1/2m$ is shorter than the Hubble radius $r_{\rm H} = 1/H$, and the dS radiation is determined by the Gibbons-Hawking temperature, regardless of spacetime dimensions, as
\begin{eqnarray}
N_{\rm GH} = e^{- \frac{m}{T_{\rm GH}}}, \quad T_{\rm GH} = \frac{H}{2\pi}. \label{GH}
\end{eqnarray}
The dS radiation (\ref{GH}) is the leading term of the exact Gibbons-Hawking radiation which has the Bose-Einstein distribution, as shown section 3.
On the other hand, in the (1+1)-dimensional Minkowski space, the Schwinger pair-production rate may be given by the Unruh temperature for the accelerating charge
\begin{eqnarray}
N_{\rm S} = e^{- \frac{m}{2 T_{\rm U}}}, \quad T_{\rm U} = \frac{qE/m}{2\pi}. \label{Sch}
\end{eqnarray}
A subtle issue in interpreting the Schwinger effect via the Unruh temperature is that the temperature measured by a detector is $T_{\rm U}$ \cite{BPS,BMPS} and
the worldline instanton path in the Eucliean proper-time gives the periodicity $1/T_{\rm U}$ \cite{Dunne-Schubert} or that another Unruh temperature $2T_{\rm U}$ for the reduced mass of the pair may give a correspondence between Schwinger mechanism and the Unruh effect \cite{CKS}. Still another interpretation is that the Schwinger effect in the (1+3)-dimensional spacetime has the Boltzmann distribution for the transverse energy with the Unruh temperature (\ref{Sch}) together with a chemical potential \cite{Hwang-Kim} or that a nonstandard gyromagnetic ratio of the charge may give the identical result from the Unruh effect and the Schwinger effect \cite{Labun-Rafelski}. The resolution of this Unruh temperature for the Schwinger mechanism is not the main issue and goes the scope of this paper.

Under the influence of the electric field, the expansion of ${\rm dS}_2$ space has the effect of separating the charged pairs and thus reducing the mass by the effective mass $\bar{m} = \sqrt{m^2 - H^2/4}$ for the pair production while the confining nature of ${\rm AdS}_2$ has the effect of binding the pair and increasing the mass by another effective mass $\bar{m} = \sqrt{m^2 + K^2/4}$. The Schwinger pair production in the ${\rm AdS}_2$ space has the Breitenlohlner-Freedman bound $qE \geq K \bar{m}$.
Now, a question is whether the Schwinger mechanism (\ref{pair}) may be interpreted in terms of an effective temperature associated with the Unruh temperature (\ref{Sch}) and the Gibbons-Hawking temperature (\ref{GH}). A naive interpretation for ${\rm dS}_2$ would be that eq. (\ref{pair}) may have the form $N_{\rm S} = e^{- (\epsilon - \mu)/T_{\rm GH}}$, where $\epsilon$ is the energy $\gamma_{dS}$ of the charge and $\mu$ is the chemical potential $qE/H$ on the horizon. A drawback of this interpretation, however, is that it does not have an analog in ${\rm AdS}_2$ because ${\rm AdS}_2$ does not have a horizon and the associated temperature.

Now, we advance a new thermal interpretation of the pair production in the uniform electric field and in ${\rm dS}_2$ and ${\rm AdS}_2$, which writes the the pair-production rate as
\begin{eqnarray}
N = e^{-\frac{\bar{m}}{T_{\rm eff}} },
\end{eqnarray}
where the effective temperature for ${\rm dS}_2$ is
\begin{eqnarray}
T_{\rm dS} = \sqrt{T_{\rm GH}^2 + T_{\rm U}^2} + T_{\rm U}, \label{ds tem}
\end{eqnarray}
and for ${\rm AdS}_2$ is
\begin{eqnarray}
T_{\rm AdS} = T_{\rm U} + \sqrt{T_{\rm U}^2 - \Bigl( \frac{K}{2 \pi} \Bigr)^2}. \label{ads tem}
\end{eqnarray}
The Breitenlohler-Freedman bound for the instability of ${\rm AdS}_2$ becomes $T_{U} \geq K/(2 \pi)$.
It is interesting to compare eqs. (\ref{ds tem}) and (\ref{ads tem}) with the effective temperature for an accelerating observer in ${\rm dS}_2$ and ${\rm AdS}_2$, which is given by \cite{NPT,Deser-Levin97}
\begin{eqnarray}
T_{\rm eff} = \sqrt{T_{\rm U}^2 + \frac{{\cal R}}{8 \pi^2}}, \label{acc tem}
\end{eqnarray}
where ${\cal R}$ denotes the scalar curvature in ${\rm dS}_2$ and ${\rm AdS}_2$. Therefore, the Schwinger mechanism in ${\rm dS}_2$ and ${\rm AdS}_2$ may have the interpretation of the temperature for an accelerating charge up to the factor of two, which is intrinsic to the Schwinger mechanism. In other words,
eqs. (\ref{ds tem}) and (\ref{ads tem}) are the effective temperatures modified by the uniform electric field. In the following sections we explore the quantum field theory for the Schwinger mechanism in ${\rm dS}_2$ and ${\rm AdS}_2$ and find the one-loop effective actions in the proper-time integral.

\section{One-Loop Effective Action in ${\rm dS}_2$}

We canonically quantize the complex field in terms of the annihilation operator for a particle and the creation operator for an antiparticle as
\begin{eqnarray}
\hat{\phi} (t,x) = \int \frac{dk}{2 \pi} \bigl[ \hat{a}_k \phi_{k}^{(+)} (t) e^{i kx} + \hat{b}^{\dagger}_k \phi_{k}^{(-)}(t) e^{- i kx} \bigr].
\end{eqnarray}
Here, the positive frequency solution $\phi_{k}^{(+)}$ and the negative frequency solution $\phi_{k}^{(-)}$ of the mode equation in ${\rm dS}_2$
\begin{eqnarray}
\ddot{\phi}_{k} + H \dot{\phi}_{k} + \omega_{k}^2 (t) \phi_{k} = 0,
\end{eqnarray}
with
\begin{eqnarray}
\omega_{k}^2 (t) = m^2 + e^{-2Ht} \Bigl(\bar{k} + \frac{qE}{H} e^{Ht} \Bigr)^2
\end{eqnarray}
satisfy the Wronskian condition
\begin{eqnarray}
e^{Ht} {\rm Wr}_{(t)} [\phi_{k}^{(+)} (t), \phi_{k}^{(-)} (t)] = i. \label{wr con}
\end{eqnarray}
The Wronskian condition (\ref{wr con}) comes from the quantization rule in curved spacetimes.

\subsection{Bogoliubov Transformation and Exact Pair-Production Rate}

The out-vacuum is defined by the positive and the negative frequency solutions in the future infinity ($t = \infty$)
\begin{eqnarray}
\phi_{{\rm out}, k}^{(+)} &=& \frac{e^{i \pi/4} e^{\pi |\mu|/2}}{\sqrt{2 \gamma}} \Bigl(\frac{H}{2 \bar{k}} \Bigr)^{3/2} e^{-Ht/2} z M_{\lambda, \mu} (z), \nonumber\\
\phi_{{\rm out}, k}^{(-)} &=& \frac{e^{i \pi/4} e^{\pi |\mu|/2}}{\sqrt{2 \gamma}} \Bigl(\frac{H}{2 \bar{k}} \Bigr)^{3/2} e^{-Ht/2} z M_{\lambda, - \mu} (z),
\end{eqnarray}
where $M_{\lambda, \mu}$ is the Whittaker function and
\begin{eqnarray}
z = 2i \frac{\bar{k}}{H} e^{-Ht}, \quad \lambda = - i \frac{qE}{H^2} \frac{\bar{k}}{|\bar{k}|}, \quad \mu = i \frac{\gamma_{dS}}{H}.
\end{eqnarray}
The in-vacuum is constructed by the positive and the negative frequency solutions in the past infinity ($t = - \infty$)
\begin{eqnarray}
\phi_{{\rm in}, k}^{(+)} &=& \frac{e^{- \pi |\lambda|/2}}{\sqrt{H}} \Bigl(\frac{H}{2 \bar{k}} \Bigr)^{3/2} e^{-Ht/2} (-z) W_{-\lambda, \mu} (-z), \nonumber\\
\phi_{{\rm in}, k}^{(-)} &=& \frac{e^{- \pi |\lambda|/2}}{\sqrt{H}} \Bigl(\frac{H}{2 \bar{k}} \Bigr)^{3/2} e^{-Ht/2} z W_{\lambda, \mu} (z),
\end{eqnarray}
where $W_{\lambda, \mu}$ is another Whittaker function and the Riemann sheet is chosen such that $- \pi \leq {\rm arg} (z) < \pi$ so that $-z = e^{-i \pi}z$. The solutions in the conformal time may be used \cite{FGKSSTV}. In this paper, we consider the case of $\bar{k} \geq 0$, but for $\bar{k} < 0$, $\lambda$ is replaced by $-\lambda$.
Using the connection formula 9.233-2 of ref. \cite{GR-table}, we find the Bogoliubov coefficients
\begin{eqnarray}
\phi_{{\rm in}, k}^{(+)} = \alpha_{k} \phi_{{\rm out}, {k}}^{(+)} + \beta_{k} \phi_{{\rm out}, {k}}^{(-)}
\end{eqnarray}
where
\begin{eqnarray}
\alpha_{k} &=& - e^{i \pi/4} \sqrt{\frac{H}{2 \gamma_{dS} }} e^{({\cal S}_{\mu}+{\cal S}_{\lambda})/4} e^{-i \pi \lambda} \frac{\Gamma(1+ 2 \mu)}{\Gamma (\frac{1}{2} + \mu - \lambda)}, \nonumber\\
\beta_{k} &=&  e^{i \pi/4} \sqrt{\frac{H}{2 \gamma_{dS} }} e^{({\cal S}_{\mu}+{\cal S}_{\lambda})/4} e^{-i \pi (\lambda - \mu- \frac{1}{2}) } \frac{\Gamma(1+ 2 \mu)}{\Gamma (\frac{1}{2} + \mu + \lambda)}.
\end{eqnarray}
Here, the instanton actions are
\begin{eqnarray}
{\cal S}_{\mu} = -2 i \pi \mu = 2 \pi \frac{\gamma_{dS}}{H}, \quad  {\cal S}_{\lambda} = 2 \pi \vert \lambda \vert = 2 \pi \frac{qE}{H^2}.
\end{eqnarray}
The Bogoliubov relation $\vert \alpha_{k} \vert^2  - \vert \beta_k \vert^2 =1$ holds.  The mean number of produced pairs is
\begin{eqnarray}
N_{\rm dS} = \vert \beta_{k} \vert^2 = \frac{e^{- ({\cal S}_{\mu}- {\cal S}_{\lambda})} + e^{- 2 {\cal S}_{\mu}}}{1-e^{- 2 {\cal S}_{\mu}}}. \label{ds ex-pair}
\end{eqnarray}
The pair-production rate (\ref{pair}) from the phase-integral method is the leading Boltzmann factor of the exact one (\ref{ds ex-pair}).

The effective mass $\bar{m} = \sqrt{m^2 - (H/2)^2}$ can be arbitrarily small when the Hubble radius is comparable to the Compton wavelength of the charge, and thus lowers the effective critical strength $E_{\rm C} = \bar{m}^2/q$ and enhances the pair-production rate as shown in the left panel of figure 1. As the Hubble radius increases, the pair-production rate decreases more rapidly for a weak field than for a strong field, but in general the pair production increases as the field gets stronger. As shown in the right panel of figure 1, the Boltzmann factor for the pair production in section 2 is a good approximation to the exact formula regardless of the field strength except when the Hubble radius is comparable to the effective Compton wavelength.
 \begin{figure}[t]
{\includegraphics[width=0.475\linewidth]{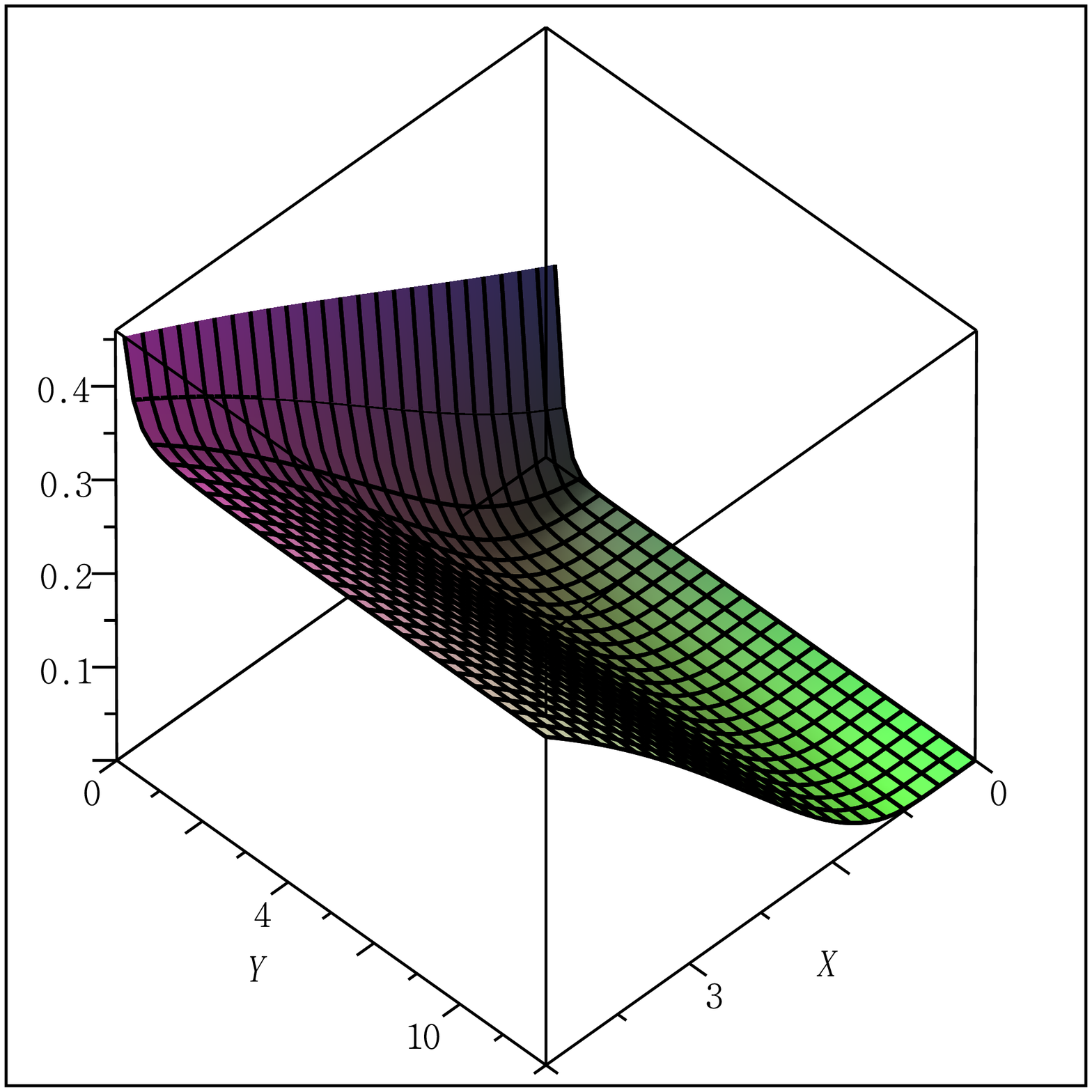}}\hfill
{\includegraphics[width=0.475\linewidth]{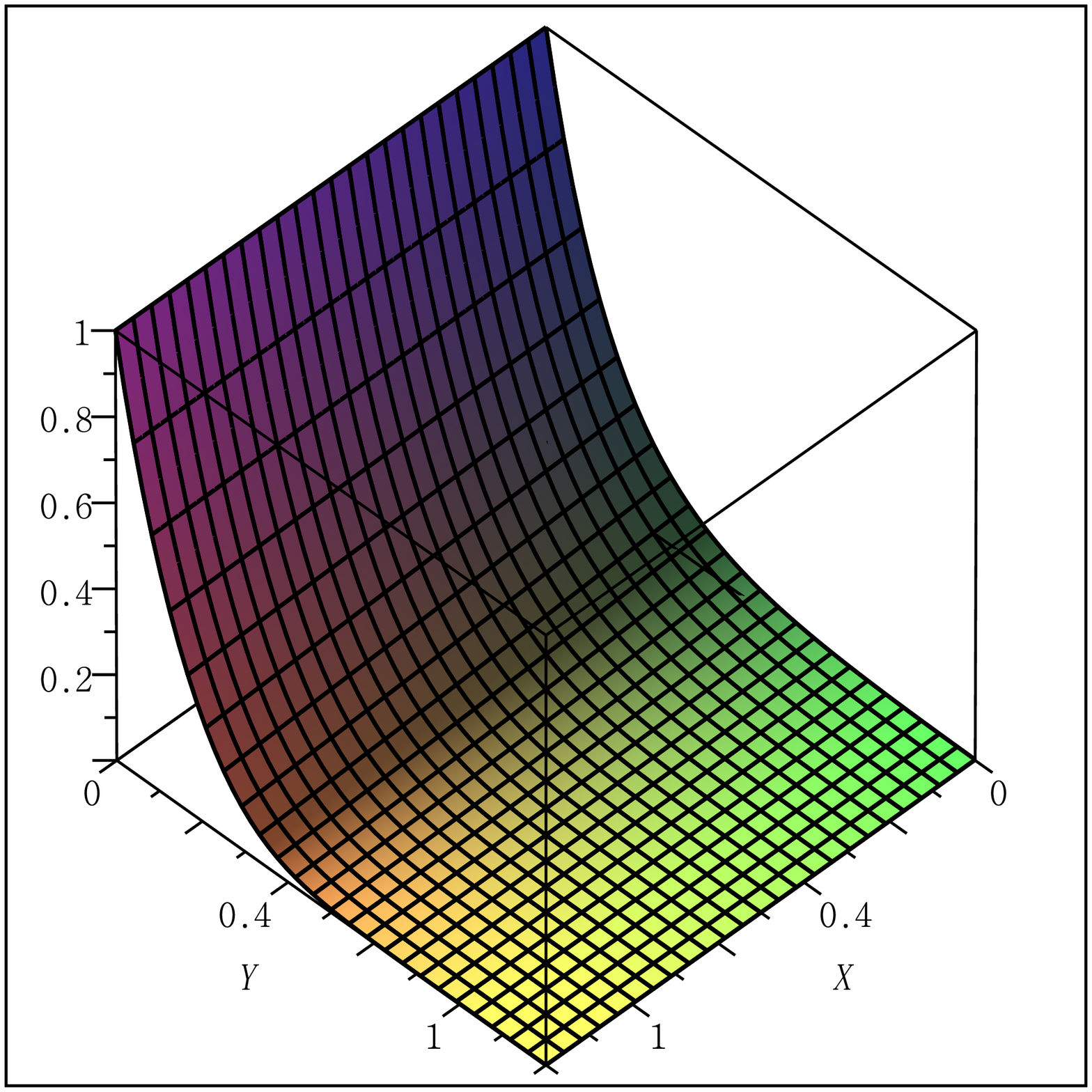}}
\caption{(color online). The Schwinger pair-production rate  $N_{\rm dS}$ in ${\rm dS}_2$ as a function of $E$ and $H$ is plotted in the range of $X = [0,3]$ and $Y = [0,10]$, where $X = \frac{qE}{\bar{m}^2} = \frac{qE}{m\sqrt{1- (H/2m)^2}}$ and $Y = \frac{\bar{m}}{H} = \sqrt{(m/H)^2 -1/4}$ [left panel], and the relative ratio of the difference between $N_{\rm dS}$ and the Boltzmann factor $N_{\rm S} = e^{- ({\cal S}_{\mu} - {\cal S}_{\lambda} )}$, that is, $1- \frac{N_{\rm S}}{N_{\rm dS}}$ for the range of $X = [0,1]$ and $Y = [0,1]$ [right panel].} \label{dS-pair plot}
\end{figure}

\subsection{Effective Action}

In the in-out formalism based on the variational principle \cite{DeWitt03,DeWitt75}, the one-loop effective action per Hubble time and per unit length is given by
\begin{eqnarray}
{\cal L}^{(1)}_{\rm dS} = i H \int \frac{dk}{2 \pi}  \ln (\alpha_{k}^*),
\end{eqnarray}
which, up to trivial terms to be regulated away, leads to
\begin{eqnarray}
{\cal L}^{(1)}_{\rm dS} = i H \int \frac{dk}{2 \pi} \Bigl[ \ln \Gamma \Bigl(1- i \frac{{\cal S}_{\mu}}{\pi} \Bigr) - \ln \Gamma \Bigl(\frac{1}{2}- i \frac{{\cal S}_{\mu}}{2\pi} + i \frac{{\cal S}_{\lambda}}{2 \pi} \Bigr)\Bigr].
\end{eqnarray}
We now employ the gamma-function regularization method in refs. \cite{Kim-Lee-Yoon08,Kim-Lee-Yoon10}, which first expresses the logarithm of the gamma function via the formula 8.341-3 of ref. \cite{GR-table} as
\begin{eqnarray}
\ln (\Gamma (z)) = \int^{\infty}_{0} \frac{ds}{s} \Bigl[ \frac{e^{-z s}}{1- e^{-s}} - \cdots \Bigr],
\end{eqnarray}
where the dots denote the subtraction of the terms to be regularized away through renormalization of the vacuum energy and coupling constants.
We then applies the Cauchy's residue theorem $\oint f(z) dz = \pm 2 \pi i \sum ({\rm residue})$, which performs a contour integral of the quarter-circle of infinite radius in the first or the fourth quadrant of the complex plane. For a complex argument of the gamma function, we obtain the complex effective action, which is the case of QED action in electric fields. Furthermore, from the Cauchy theorem does follow the consistence relation between the imaginary part of the effective action and the pair-production rate:
\begin{eqnarray}
2 {\rm Im} {\cal L}^{(1)} = \sum \ln (\vert \alpha_k \vert^2 ) = \sum \ln ( 1 + \vert \beta_k \vert^2).
\end{eqnarray}
Finally, we find the vacuum polarization (real part of the effective action) in the proper-time integral
\begin{eqnarray}
{\cal L}^{(1)}_{\rm dS} &=& \frac{qE}{2 \pi} \int_{0}^{\infty} \frac{ds}{s} \Bigl[ e^{ - ({\cal S}_{\mu} - {\cal S}_{\lambda})s/2\pi} \Bigl(\frac{1}{\sin(s/2)} - \frac{2}{s} - \frac{s}{12} \Bigr) \nonumber\\ &&- e^{ - {\cal S}_{\mu}s/\pi} \Bigl(\frac{\cos(s/2)}{\sin(s/2)} - \frac{2}{s} +\frac{s}{6} \Bigr) \Bigr]. \label{ds action}
\end{eqnarray}
In the above, we have used the Schwinger substraction scheme, in which the first subtracted term renormalizes the vacuum energy and the second one
renormalizes the charge in the large-field limit, though the charge renormalization involves a finite term \cite{Blau-Visser-Wipf} in 1+1 dimensions. In obtaining eq. (\ref{ds action}) we have counted the number of states of longitudinal momentum and have used the symmetric argument.

A few comments are in order. Firstly, summing all the principal values of simple poles located on the positive imaginary axis, we obtain the imaginary part of the effective action
\begin{eqnarray}
{\rm Im} {\cal L}^{(1)}_{\rm dS} = \frac{qE}{2(2 \pi)} \bigl[\ln (1+ e^{-({\cal S}_{\mu} - {\cal S}_{\lambda})}) - \ln (1- e^{-2 {\cal S}_{\mu}}) \bigr]. \label{ds vac-per}
\end{eqnarray}
The vacuum persistence is identical with that from the principal values of the effective action (\ref{ds action}). Furthermore, the vacuum persistence satisfies
the consistency relation for the scalar QED
\begin{eqnarray}
2 {\rm Im} {\cal L}^{(1)}_{\rm dS} =  \frac{qE}{2 \pi} \ln ( 1 + N_{\rm dS} ).
\end{eqnarray}
Secondly, the first integral of eq. (\ref{ds action}) corresponds to the scalar QED action while the second integral corresponds to the spinor QED action in a constant electric field in Minkowski spacetime \cite{Schwinger}. It is the consequence between the vacuum persistence and the distribution function of produced pairs: the Bose-Einstein distribution gives the vacuum persistence for the spinor QED while the Fermi-Dirac distribution gives that for the scalar QED \cite{Stephens}. The effective action for the pure dS space without the electric field is obtained by combining both integrals while QED action in the Minkowski space ($H = 0$) is given by the first integral because ${\cal S}_{\mu} = \infty$. Finally, a non-minimal coupling of the field $\xi {\cal R} \phi^* \phi$ can change the effective mass by $\bar{m} = \sqrt{m^2 + (\xi- 1/8) {\cal R}}$. In this paper we consider only QED with the minimal coupling with the abelian gauge for electromagnetic fields.

We now investigate the effect of the dS radiation on the Schwinger effect and, conversely, the effect of the Schwinger effect on the dS radiation, that is, the curvature effect on the Schwinger mechanism and the Schwinger effect on the spacetime curvature. For that purpose, we express the effective action in terms of the rescaled scalar curvature and the Maxwell scalar
\begin{eqnarray}
\bar{\cal R}= \frac{1}{2} {\cal R} = H^2, \quad \bar{\cal F} = - 2 q^2 {\cal F} = (qE)^2.
\end{eqnarray}
The effective action (\ref{ds action}) has the power series expansion \cite{GR-table}
\begin{eqnarray}
{\cal L}^{(1)}_{\rm dS} = \frac{qE}{2 \pi} \sum_{n=2}^{\infty} \frac{(-1)^{n-1} 2 B_{2n}}{(2n) (2n-1)} \Bigl[ (1-2^{1-2n} ) \Bigl(\frac{2 \pi}{{\cal S}^{(-)} } \Bigr)^{2n-1} - \Bigl(\frac{2 \pi}{ {\cal S}_{\mu} } \Bigr)^{2n-1}  \Bigr], \label{ds ser}
\end{eqnarray}
where $B_{2n}$ are Bernoulli numbers, and ${\cal S}^{(-)}= {\cal S}_{\mu}- {\cal S}_{\lambda}$ and ${\cal S}_{\mu}$ are given by
\begin{eqnarray}
{\cal S}^{(-)} &=& 2 \pi \sqrt{ \frac{\bar{m}^2}{ \bar{\cal R} } + \frac{\bar{\cal F}}{ \bar{\cal R}^2}  } - 2 \pi \frac{ \sqrt{ \bar{\cal F}} }{ \bar{\cal R}}, \nonumber\\
{\cal S}_{\mu} &=& 2 \pi \sqrt{ \frac{\bar{m}^2}{ \bar{\cal R} } + \frac{\bar{\cal F}}{ \bar{\cal R}^2 }  }. \label{ds S}
\end{eqnarray}
The first term of (\ref{ds ser}) gives
\begin{eqnarray}
{\cal L}^{(1)}_{\rm dS} = \frac{1}{360 \pi} \frac{\sqrt{\bar{\cal F}}}{\bar{m}^6}  \Biggl[ \frac{7}{8}
 \Bigl(\sqrt{ \bar{\cal F} + \bar{m}^2 \bar{\cal R} } + \sqrt{\bar{\cal F} } \Bigr)^3 - \Bigl( \frac{\bar{m}^4 \bar{\cal R}^2}{ \bar{\cal F} + \bar{m}^2 \bar{\cal R} } \Bigr)^{3/2}  \Biggr]. \label{ds 1-term}
\end{eqnarray}
In the limit of the zero-scalar curvature, the first term (\ref{ds 1-term}) reduces to the Schwinger action in the two-dimensional Minkowski spacetime.
It will be interesting to compare the non-power expansion of the Maxwell scalar and scalar curvature in eqs. (\ref{ds ser}) and (\ref{ds 1-term}) with some perturbative power expansions of the one-loop effective action for the Einstein-Maxwell theory \cite{Avramidi-Fucci,Davila-Schubert}.

In the strong-field limit ($\sqrt{\bar{\cal F}} \gg \bar{\cal R}$), keeping the first two leading terms of eq. (\ref{ds S}), the effective action  approximately takes the form
\begin{eqnarray}
{\cal L}^{(1)}_{\rm dS} &=& \frac{qE}{2 \pi} \int_{0}^{\infty} \frac{ds}{s} e^{ - (\bar{m}^2/ 2 qE)s}
\Bigl( e^{(\bar{m}^4 \bar{\cal R}^3/\bar{\cal F}^{3/2}) s} - 2 e^{- (\sqrt{\bar{\cal F}}/\bar{\cal R})s} \cos^2 (s/4) \Bigr) \nonumber \\ && \times \Bigl(\frac{1}{\sin(s/2)} - \frac{2}{s} - \frac{s}{12} \Bigr). \label{ds st-E}
\end{eqnarray}
Note that the first and the last factor is the one-loop effective action in the Minkowski spacetime whereas the middle parenthesis is the leading curvature effect on the Schwinger effect. In the limit of the zero-scalar curvature, the expansion reduces to the first term of the Schwinger action in the two-dimensional Minkowski spacetime
\begin{eqnarray}
{\cal L}^{(1)}_{\rm dS} = \frac{7}{720 \pi} \frac{(qE)^4}{m^6}.
\end{eqnarray}
In the opposite case of the strong gravity limit ($\bar{\cal R} \gg \sqrt{\bar{\cal F}}$), the effective action is approximately given by
\begin{eqnarray}
{\cal L}^{(1)}_{\rm dS} &=& \frac{qE}{2 \pi} \int_{0}^{\infty} \frac{ds}{s} e^{ - (\bar{m}/ H)s}  e^{- (\bar{\cal F}/\bar{m} \bar{\cal R}^{3/2})s}
\Bigl( \frac{ e^{(\sqrt{\bar{\cal F}}/\bar{\cal R})s} -1 }{\cos(s/2)} -1 \Bigl) \nonumber\\&& \times \Bigl(\frac{\cos(s/2)}{\sin(s/2)} - \frac{2}{s} + \frac{s}{6} \Bigr). \label{ds st-R}
\end{eqnarray}
In the extreme case of the zero field ($\bar{\cal F} = 0$), the number of states $qE/(2 \pi)$ should be replaced by $H^2$ per unit Hubble length and per unit Hubble time. Then, the one-loop effective action takes the form of spinor QED action, where $\bar{m}^2/(2qE)$ is replaced by $\bar{m}/H$ in the exponent and $qE/(2 \pi)$ by $H^2$ in the prefactor. This apparent spin-statistics may be understood from the fact that the dS radiation, a Bose-Einstein distribution, has the vacuum persistence for the spinor QED \cite{Stephens}
\begin{eqnarray}
{\rm Im} {\cal L}^{(1)}_{\rm dS} = - \frac{H^2}{2} \ln (1 - e^{-\bar{m}/T_{\rm GH} }). \label{ds rad}
\end{eqnarray}
Because the vacuum persistence is determined by the poles of the one-effective action in the proper-time integral, the effective action also should have the form for the opposite spin-statistics, the last parenthesis in Eq. (\ref{ds st-R}).

Interestingly, the vacuum persistence (\ref{ds vac-per}) and the exact pair-production rate (\ref{ds ex-pair}) may be written in terms of the effective temperature in section 2 and an additional one
\begin{eqnarray}
{\cal S}^{(-)} = \frac{\bar{m}}{T_{\rm dS}}, \quad {\cal S}_{\mu} = \frac{\bar{m}}{\bar{T}_{\rm dS}},
\end{eqnarray}
where
\begin{eqnarray}
T_{\rm dS} = \sqrt{T_{\rm U}^2 + T_{\rm GH}^2} + T_{\rm U}, \quad \bar{T}_{\rm dS} = \frac{T_{\rm GH}^2}{2 \sqrt{T_{\rm U}^2+T_{\rm GH}^2}}.
\end{eqnarray}
The effective temperature $T_{\rm dS}$ has the interpretation of the accelerating charge in ${\rm dS}_2$ but $\bar{T}_{\rm dS}$ does not have any gravity analog.
In the case of strong field, in which $T_{\rm dS} \approx 2 T_{\rm U}$ and $\bar{T}_{\rm dS} \approx  T^2_{\rm GH}/(2T_{\rm U})$, the exact pair-production rate (\ref{ds ex-pair}) is dominated by the Schwinger mechanism and the corrections are exponentially suppressed. In the opposite case of strong gravity, in which $T_{\rm dS} \approx T_{\rm GH}$ and $\bar{T}_{\rm dS} \approx  T_{\rm GH}/2$, the Schwinger term and the correction terms are comparable each other, which can be summed as the Bose-Einstein distribution of the dS radiation.

\section{One-Loop Effective Action in ${\rm AdS}_2$}

The universal covering space of AdS space obtained by unwrapping $S^1$ will be used to quantize the quantum field so that there is no closed timelike curve in the metric (\ref{ads gauge}). The energy mode $\phi = e^{- i \omega t} \varphi_{\omega} (x)$ of the charged scalar field satisfies the equation
\begin{eqnarray}
\Bigl[e^{-Kx} \partial_x (e^{Kx} \partial_x) + e^{-2Kx} \Bigl(\bar{\omega} - \frac{qE}{K} e^{Kx} \Bigr)^2 - m^2 \Bigr] \phi_{\omega} (x) = 0.
\end{eqnarray}
The energy can take any value as in the Minkowski spacetime because the topology $S^1$ for time coordinate is unwrapped. The quantum field will be considered for the pair production so that the Breitenlohlner-Freedman bound for the instability holds. As for ${\rm dS}_2$, at the asymptotic boundary $x = \infty$, we find the solutions with the positive (outgoing) flux and the negative (incoming) flux, respectively,
\begin{eqnarray}
\phi_{\omega}^{(+)} (x) &=& \Bigl( \frac{K}{2i \bar{\omega}} \Bigr)^{1/2} \Bigl( \frac{1}{2 \gamma_{AdS}} \Bigr)^{1/2} M_{\kappa, - \nu} (\zeta), \nonumber\\
\phi_{\omega}^{(-)} (x) &=& \Bigl( \frac{K}{2i \bar{\omega}} \Bigr)^{1/2} \Bigl( \frac{1}{2 \gamma_{AdS}} \Bigr)^{1/2} M_{\kappa, \nu} (\zeta), \label{ads sol}
\end{eqnarray}
where
\begin{eqnarray}
\kappa = i \frac{qE}{K^2}, \quad \nu = i \frac{\gamma_{AdS}}{K}, \quad \zeta = 2 i \frac{\bar{\omega}}{K} e^{-Kx}.
\end{eqnarray}
The solutions (\ref{ads sol}) satisfy the quantization condition for the tunneling boundary for QED in static electric fields \cite{Kim-Lee-Yoon10}
\begin{eqnarray}
e^{Kx} {\rm Wr}_{(x)} [ \phi^{(-)} (x), \phi^{(+)} (x)  ] = i. \label{ads wr}
\end{eqnarray}
Similarly, at $x = - \infty$ the other set of solutions with the positive (incoming) flux and the negative (outgoing) flux satisfying the condition  (\ref{ads wr}) are
\begin{eqnarray}
\phi_{\omega}^{(+)} (x) &=& \Bigl( \frac{1}{2 \bar{\omega}} \Bigr)^{1/2} W_{\kappa, - \nu} (\zeta), \nonumber\\
\phi_{\omega}^{(-)} (x) &=& \Bigl( \frac{1}{2 \bar{\omega}} \Bigr)^{1/2} W_{- \kappa, - \nu} (e^{-i \pi} \zeta).
\end{eqnarray}

Using the connection formula 9.233-2 between the Whittaker functions \cite{GR-table}
\begin{eqnarray}
M_{\kappa, - \nu} (\zeta) &=& \frac{\Gamma (1 - 2 \nu)}{\Gamma (\frac{1}{2} - \nu - \kappa)} e^{- i \pi \kappa}
  W_{- \kappa, - \nu} (e^{-i \pi} \zeta) \nonumber\\&& + \frac{\Gamma (1 - 2 \nu)}{\Gamma (\frac{1}{2} - \nu + \kappa)} e^{- i \pi (\kappa+ \nu - \frac{1}{2})} W_{\kappa, - \nu} (\zeta),
\end{eqnarray}
we find the Bogoliubov coefficient as the ratio of the coefficient for the outgoing flux solution to the coefficient of the incoming flux solution, which is given by
\begin{eqnarray}
\alpha_{\omega} =  \frac{\Gamma (\frac{1}{2} - \nu + \kappa)}{\Gamma (\frac{1}{2} - \nu - \kappa)} e^{- i \pi \nu}.
\end{eqnarray}
The one-loop effective action in the in-out formalism is obtained from the contour integrals of quarter circles in the first and fourth quadrants as
\begin{eqnarray}
{\cal L}^{(1)}_{\rm AdS} = - \frac{qE}{2 \pi} \int^{\infty}_{0} \frac{ds}{s} e^{- {\cal S}_{\kappa} s/(2 \pi)} \cosh({\cal S}_{\nu} s/2 \pi) \Bigl[ \frac{1}{\sin(s/2)} - \frac{2}{s} - \frac{s}{12} \Bigr], \label{ads act}
\end{eqnarray}
where
\begin{eqnarray}
{\cal S}_{\kappa} =- 2 \pi i \kappa = 2 \pi \frac{qE}{K^2}, \quad {\cal S}_{\nu} =- 2 \pi i \nu = 2 \pi \frac{\gamma_{AdS}}{K}.
\end{eqnarray}
Here, we have used the Schwinger subtraction scheme by subtracting the first term for the vacuum energy renormalization and the second term for the finite charge renormalization. The contour integrals give the vacuum persistence
\begin{eqnarray}
2 ~ {\rm Im} ({\cal L}^{(1)}_{\rm AdS}) = \ln \Bigl (  \frac{1 +  e^{-({\cal S}_{\kappa} -  {\cal S}_{\nu})} }{1+ e^{-({\cal S}_{\kappa} +  {\cal S}_{\nu})}} \Bigr) = \ln (1+ N_{\rm AdS} ). \label{ads vac-per}
\end{eqnarray}
The vacuum persistence can also be obtained from the principle values of the effective action (\ref{ads act}).
The vacuum persistence is consistent with the number of produced pairs
\begin{eqnarray}
N_{\rm AdS} = \vert \alpha_{\omega} \vert^2 -1 = \frac{e^{-({\cal S}_{\kappa} -  {\cal S}_{\nu})} - e^{-({\cal S}_{\kappa} +  {\cal S}_{\nu})} }{1+ e^{-({\cal S}_{\kappa} +  {\cal S}_{\nu})}}. \label{ads pair}
\end{eqnarray}
Note that the near-horizon geometry of an extremal RN black hole is ${\rm AdS}_2$ with a uniform electric field and that the pair-production rate (\ref{ads pair}) is the same as the emission rate of charged $s$-wave from the extremal RN black hole with the charge $Q$, where the electric field is $E = 1/Q$ and the curvature is $1/Q^2$ \cite{CKLSW}. The pair production in ${\rm AdS}_2$ thus may shed light on understanding the emission from RN black holes beyond the leading Schwinger term \cite{Gibbons}.

The leading term $N_{\rm S} = e^{-({\cal S}_{\kappa} -  {\cal S}_{\nu})}$ is the Boltzmann factor for the Schwinger formula in section 2. The binding nature of ${\rm AdS}_2$ provides the effective mass $\bar{m} = m\sqrt{1+ (K/2m)^2}$ and thus increases the critical strength $E_{\rm C} = \bar{m}^2/q$. The ratio of $1/K$ to the effective Compton wavelength has the lower bound $Y = 1/2$, and the Breitenlohler-Freedman bound gives a constraint $X \geq 1/Y$ in figure 2. As shown in the left panel of figure 2 the pair production drastically increases near $Y =1/2$ as the field increases but gently increases otherwise. The right panel of figure 2 shows that the Boltzmann factor is a good approximation except for the region near $Y = 1/2$, where the curvature effect becomes important.
\begin{figure}[t]
{\includegraphics[width=0.475\linewidth]{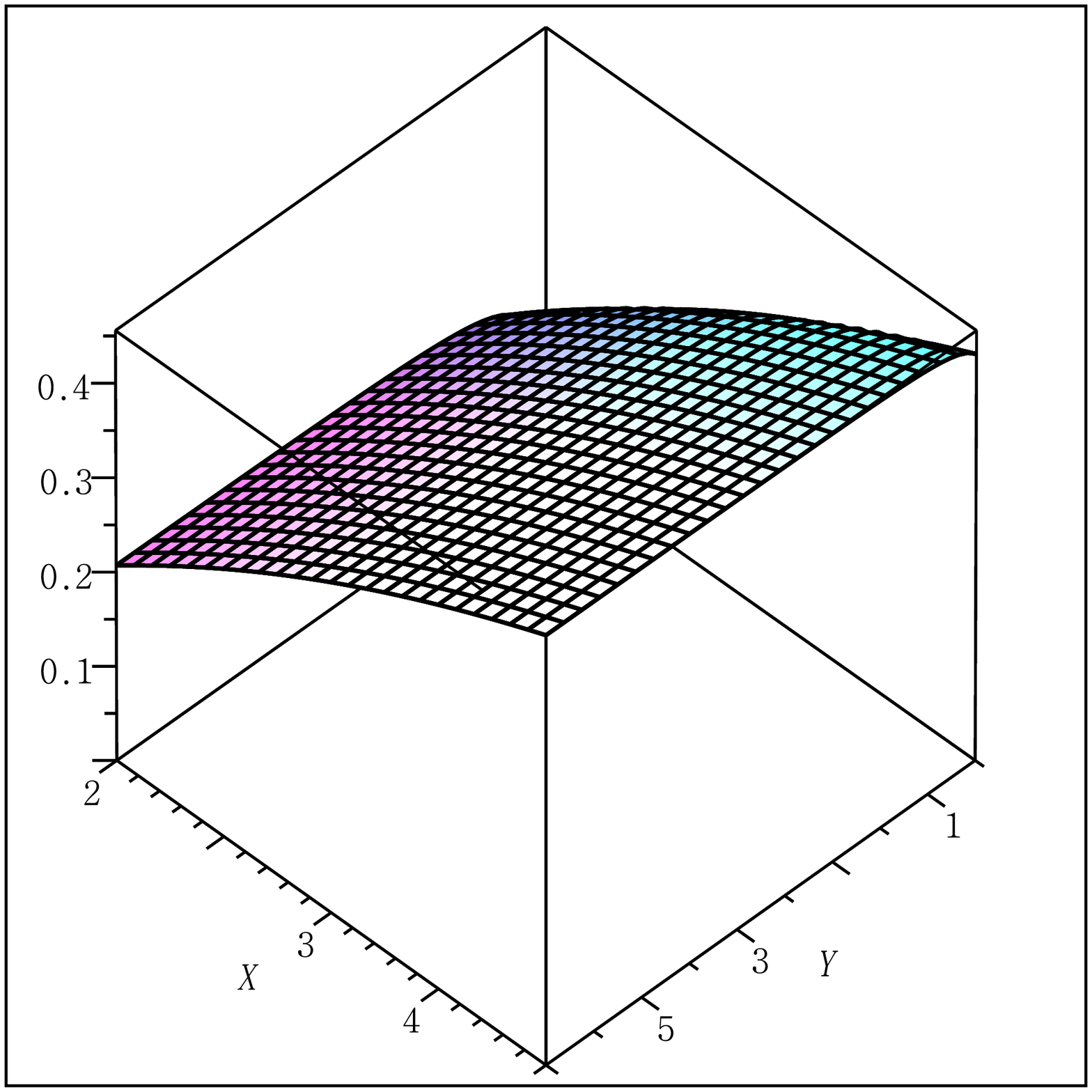}}\hfill
{\includegraphics[width=0.475\linewidth]{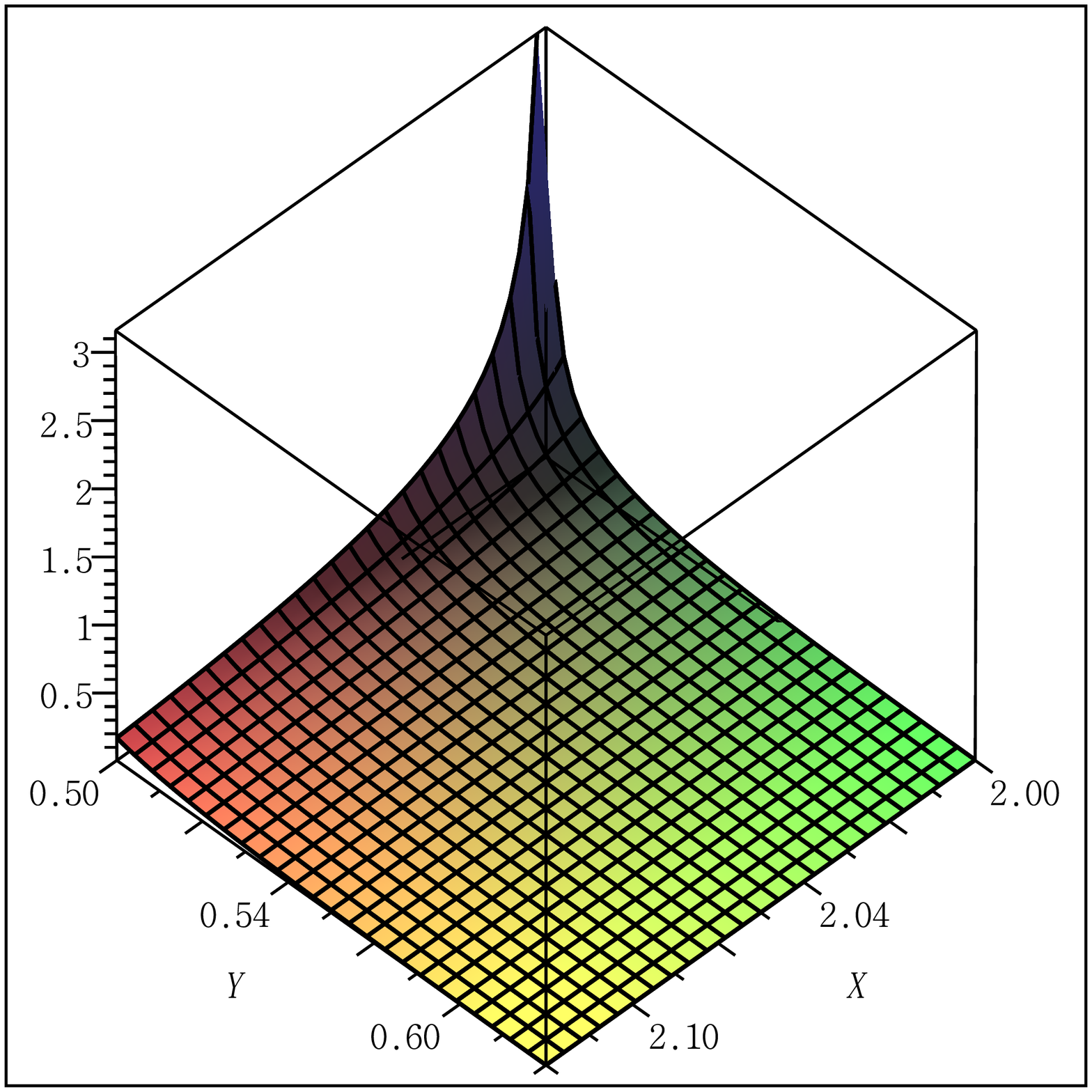}}
\caption{(color online). The Schwinger pair-production rate  $N_{\rm AdS}$ in ${\rm AdS}_2$ as a function of $E$ and $K$ is plotted in the range of $X = [2,4]$ and $Y = [\frac{1}{2},5]$, where $X = \frac{qE}{\bar{m}^2}=\frac{qE}{m\sqrt{1+(K/2m)^2}}$ and $Y = \frac{\bar{m}}{K} = \sqrt{\frac{m^2}{K^2} + \frac{1}{4}}$ [left panel], and the relative ratio of the difference between Boltzmann factor $N_{\rm S} = e^{- ({\cal S}_{\kappa} - {\cal S}_{\nu} )}$ and $N_{\rm AdS}$, that is, $\frac{N_{\rm S}}{N_{\rm AdS}} -1$ for the range of $X = [2,\frac{21}{10}]$ and $Y = [\frac{1}{2},\frac{6}{10}]$ [right panel].} \label{AdS-pair plot}
\end{figure}

The one-loop effective action has a power series expansion in terms of the scalar curvature and the Maxwell scalar
\begin{eqnarray}
{\cal L}^{(1)}_{\rm AdS} = - \frac{qE}{2 \pi} \sum_{n =2}^{\infty} \frac{(-1)^{n-1} (1-2^{1-2n})B_{2n}}{(2n)(2n-1)} \Bigl[ \Bigl(\frac{2 \pi}{{\cal S}^{(-)}} \Bigr)^{2n-1} - \Bigl(\frac{2 \pi}{{\cal S}^{(+)}} \Bigr)^{2n-1} \Bigr], \label{ads ser}
\end{eqnarray}
where $B_{2n}$ are the Bernoulli numbers, and ${\cal S}^{(\mp)} ={\cal S}_{\kappa} \mp {\cal S}_{\nu}$ are given by the rescaled quantities as
\begin{eqnarray}
{\cal S}^{(\mp)} = 2  \pi \frac{ \sqrt{ \bar{\cal F} } }{ \bar{\cal R} } \mp 2 \pi \sqrt{ \frac{ \bar{\cal F} }{ \bar{\cal R}^2 } + \frac{ \bar{m}^2 }{ \bar{\cal R} } }.
\end{eqnarray}
The overall negative sign in eq. (\ref{ads act}) comes from the tunneling boundary condition, which occurs for the QED action in the Coulomb gauge \cite{Kim-Lee-Yoon10}. The expansion (\ref{ads act}) is still a non-power expansion of the Maxwell scalar and the scalar curvature, and may have a power series only in the limit of large field or curvature.

\section{Conclusion}

We have studied the Schwinger pair production by a uniform electric field in ${\rm dS}_2$ and ${\rm AdS}_2$ and obtained the exact one-loop effective actions from the scattering matrix between the in-vacuum and the out-vacuum in the in-out formalism. Furthermore, we have proposed a thermal interpretation of the pair-production rate, whose temperature is the geometric mean of the exponent two of the Gibbons-Hawking temperature and the Unruh temperature for an accelerating charge in ${\rm dS}_2$ and an analogous expression in ${\rm AdS}_2$. As far as the electromagnetic interaction is concerned, the pairs of charged particle experience an effective mass or mass gap, which decreases due to the separation of pairs in ${\rm dS}_2$, but increases due to binding of pairs in ${\rm AdS}_2$.

In ${\rm dS}_2$ the exact pair-production rate (\ref{ds ex-pair}) has the form
\begin{eqnarray}
N_{\rm dS} = \frac{e^{-\bar{m}_{\rm dS}/T_{\rm dS}} + e^{- \bar{m}_{\rm dS}/\bar{T}_{\rm dS}}}{1- e^{- \bar{m}_{\rm dS}/\bar{T}_{\rm dS}}}, \label{dS therm pair}
\end{eqnarray}
where the effective mass is $\bar{m}_{\rm dS} = m \sqrt{1 - {\cal R}_{\rm dS}/(8m^2)}$ and the effective temperatures are
\begin{eqnarray}
T_{\rm dS} = \sqrt{T_{\rm U}^2 + T_{\rm GH}^2 } + T_{\rm U}, \quad \bar{T}_{\rm dS} = \frac{T_{\rm GH}^2}{2 \sqrt{T_{\rm U}^2 + T_{\rm GH}^2}}. \label{dS temps}
\end{eqnarray}
On the other hand, in ${\rm AdS}_2$ the exact pair-production rate (\ref{ads pair}) can be analogously written as
\begin{eqnarray}
N_{\rm AdS} = \frac{e^{-\bar{m}_{\rm AdS} /T_{\rm AdS}} - e^{- \bar{m}_{\rm AdS}/\bar{T}_{\rm AdS}}}{1+ e^{- \bar{m}_{\rm AdS}/\bar{T}_{\rm AdS}}}, \label{AdS therm pair}
\end{eqnarray}
where the effective mass is $\bar{m}_{\rm AdS} = m \sqrt{1 - {\cal R}_{\rm AdS}/(8m^2)}$ and the effective temperatures are
\begin{eqnarray}
T_{\rm AdS} = T_{\rm U} + \sqrt{T_{\rm U}^2 + \frac{{\cal R}_{\rm AdS}}{8 \pi^2}}, \quad \bar{T}_{\rm AdS} = T_{\rm U} - \sqrt{T_{\rm U}^2 + \frac{{\cal R}_{\rm AdS}}{8 \pi^2}}. \label{AdS temps}
\end{eqnarray}
In the limit of the zero-scalar curvature, the formulae (\ref{dS therm pair}) and (\ref{AdS therm pair}) reduce to the Schwinger formula with $T = 2 T_{\rm U}$ in the two-dimensional Minkowski space. Interestingly, the leading Boltzmann factor for the pair-production rate respects the duality ${\cal R}_{\rm dS} \Leftrightarrow {\cal R}_{\rm AdS}$ because $ T_{\rm GH}^2 = {\cal R}_{\rm dS}/(8 \pi^2)$. However, the exact formulae do not respect this duality.

 We have found the one-loop effective actions in the in-out formalism by Schwinger and DeWitt, which expresses the quantum actions in terms of the Bogoliubov coefficients between the in-vacuum and the out-vacuum. The Bogoliubov coefficients in ${\rm dS}_2$ and ${\rm AdS}_2$ are given by gamma functions and the gamma-function regularization method leads to one-loop actions in the proper-time integral. Remarkably, the one-loop effective action in the proper-time integral is entirely determined by the Maxwell scalar and the scalar curvature.
In the zero-curvature limit, the one-loop effective action reduces to the Schwinger effective action in the Minkowski space while in the zero-field limit, the effective action action in ${\rm dS}_2$  provides the one for pure dS space. The one-loop effective action in ${\rm AdS}_2$ is found when the Breitenlohler-Freedman bound is violated and allows the pair production by the electric field. The pole structure of the one-loop effective actions correctly give the vacuum persistence and the pair-production rate in both ${\rm dS}_2$ and ${\rm AdS}_2$. We have studied the curvature effect on the Schwinger mechanism and also the Schwinger effect on the dS radiation.

\section*{Acknowledgments}

The authors would like to thank Kei-ichi Maeda and Robert Wald for helpful discussions. S.P.K. would like to appreciate the warm hospitality at Kavli Institute of Theoretical Physics, Chinese Academy of Sciences, during the KITPC Program ``Quantum Gravity, Black Holes and Strings,'' where this paper was initiated and completed.
The work of R-G.C. was supported in part by the National Natural Science Foundation of China (No.10821504, No.11035008, and No.11375247), the National Basic
Research Program of China under Grant No.2010CB833004, and by the Strategic Priority Research Program ``The Emergence of Cosmological Structures'' of the Chinese Academy of Sciences, Grant No. XDB09000000.
The work of S.P.K. was supported in part by Basic Science Research Program through the National Research Foundation of Korea (NRF) funded by the Ministry of Education (Grant No. NRF-2012R1A1B3002852).

\end{document}